\documentclass[12pt]{article}
\usepackage{amsfonts}

\usepackage{latexsym}
\usepackage{amssymb}
\usepackage{epsfig}
\usepackage{amsmath}
\usepackage{amsthm}
\usepackage[mathscr]{eucal}
\usepackage{subfigure}
\usepackage{graphicx}
\usepackage{hyperref}

\usepackage[linesnumbered,ruled,vlined, norelsize]{algorithm2e}

\makeatletter
\newcommand{\removelatexerror}{\let\@latex@error\@gobble}
\makeatother

\SetKwInOut{Input}{Input}
\SetKwInOut{Output}{Output\,}
\SetKwInOut{Data}{Data}
\SetKwProg{SeqDecode}{SeqDecode}{}{EndSeqDecode}
\SetKwProg{SeqDecodeMod}{SeqDecodeMod}{}{EndSeqDecodeMod}
\SetKwProg{VTDecode}{VTDecode}{}{EndVTDecode}
\SetKwProg{BurstDecode}{BurstDecode}{}{EndBurstDecode}
\SetKwProg{OrdDecode}{OrdDecode}{}{EndOrdDecode}
\SetKwProg{MultDecode}{MultDecode}{}{EndMultDecode}

\makeatletter
\def\BState{\State\hskip-\ALG@thistlm}
\makeatother

\newtheorem{Theorem}{Theorem}

\newtheorem{Lemma}{Lemma}[section]

\renewcommand{\qed}{\hfill{\ \ \rule{2mm}{2mm}} \vspace{0.2in}}

\newcommand{\ind}{1\hspace{-2.3mm}{1}}

\begin{document}

\title{Correcting an ordered deletion-erasure}
\author{ \textbf{Ghurumuruhan Ganesan}
\thanks{E-Mail: \texttt{gganesan82@gmail.com} } \\
\ \\
New York University, Abu Dhabi}
\date{}
\maketitle

\begin{abstract}
In this paper, we show that the single deletion correcting Varshamov-Tenengolts code, with minor modifications, can also correct an ordered deletion-erasure pattern where one deletion and at most one erasure occur and the deletion always occurs before the erasure. For large code lengths, the constructed code has the same logarithmic redundancy as optimal codes. 

\vspace{0.1in} \noindent \textbf{Key words:} Ordered deletion-erasure, redundancy.

\end{abstract}

\bigskip

\renewcommand{\theequation}{\arabic{section}.\arabic{equation}}
\setcounter{equation}{0}
\section{Introduction} \label{intro}
Consider a binary word being transmitted through a communication channel that introduces deletions and erasures in the individual bits. Levenshtein (1965)
showed that the Varshamov-Tenengolts (VT) codes introduced in Varshamov and Tenegolts (1965) can be used to correct a single deletion, no matter the location of the deleted bit within the word. Moreover, the redundancy of the code is optimal and grows logarithmically in the code length.

Since then, many aspects of deletion channels have been studied. Orlitsky (1993) obtained lower bounds for communication complexity in bidirectional exchanges for remotely located files corrupted by deletions. Helberg et al (2002) proposed codes for correcting multiple deletions but with redundancy that grows linearly with the code length. Later Brasniek et al (2016) used hashing techniques to construct low redundancy codes for correcting multiple deletions but with the caveat that the codewords belonged to a set of strings rich in certain predetermined patterns. Recently, Schoeny et al (2017) proposed a class of shifted VT codes to deal with burst deletions of fixed length. For a survey of literature on deletion channels, see Mitzenmacher (2009).

Random codes for deletions have also been studied before. From a communication complexity perspective, Orlitsky (1993) obtained bounds for file synchronization via deletion correction  with differing constraints on the number of rounds communication allowed. In a related work, Venkatramanan et al (2015) focused on developing bidirectional interactive algorithms with low information exchange. Recently, Hanna and Rouayheb (2017)proposed Guess and Check codes that map blocks of codewords to symbols in higher fields and used Reed-Solomon erasure decoding to correct words corrupted by a fixed number of deletions distributed randomly across the word.

Our analysis in this paper is more closely related to the original work of Levenshtein (1965) in that we are interested in identifying error patterns that include a single deletion and can still be corrected with logarithmic redundancy. Our main result (Theorem~\ref{thm1}) identifies one such example as an ordered deletion-erasure where a deletion and an erasure occur and the deletion always occurs before the erasure. In fact, the single deletion correcting VT codes, with minor modifications can be used to correct an ordered deletion-erasure and the construction is optimal in terms of redundancy, for large block lengths.




\subsection*{System Description}
For integer~\(n \geq 2,\) an~\(n-\)length word~\(\mathbf{x}\) is an element of~\(\{0,1\}^{n}.\) For integers~\(1 \leq d \leq e \leq n,\) the word~\(\mathbf{x}\) corrupted by an ordered deletion-erasure pattern (at positions~\(d\) and~\(e\)) is denoted as~\(F_{d,e}(\mathbf{x}) := \mathbf{y} = (y_1,\ldots,y_{n-1})\) and defined as follows. For~\(1 \leq i \leq d-1,\)~\(y_i = x_i\) and for~\(d \leq i \leq n-1, i \neq e,\)~\(y_i = x_{i+1}.\) Further if~\(e \leq n-1,\) then~\(y_e =\varepsilon,\) the erasure symbol. In words,~\(\mathbf{y}\) is obtained after deleting the bit~\(x_d\) and erasing the bit~\(x_{e+1}\) from the word~\(\mathbf{x},\) for~\(1 \leq e \leq n-1.\) If~\(e = n,\) then~\(\mathbf{y}\) is simply obtained from~\(\mathbf{x}\) after deleting the bit~\(x_d.\) Throughout we assume that~\(d \leq e;\) i.e., the deletion always occurs before the erasure (if it also occurs) and given the word~\(\mathbf{y},\) it is of interest to recover the original word~\(\mathbf{x}.\)

An~\(n-\)length code of size~\(q\) is a collection of words~\(\{\mathbf{x}_1,\ldots,\mathbf{x}_q\} \subset \{0,1\}^{n}.\) A code~\({\cal C}\) is said to be capable of correcting all ordered deletion-erasure patterns if for all~\(\mathbf{x}_1 \neq \mathbf{x}_2 \in {\cal C}\) and any~\(1 \leq d \leq e \leq n,\) we have that~\(F_{d,e}(\mathbf{x}_1) \neq F_{d,e}(\mathbf{x}_2).\) Apart from constructing codes that are capable of correcting ordered deletion-erasures, we are also interested in designing an algorithm to recover~\(\mathbf{x}\) from~\(F_{d,e}(\mathbf{x}).\)

For any~\(n-\)length code~\({\cal C},\) the redundancy of~\({\cal C}\) is defined to be
\begin{equation}\label{red_def}
R({\cal C}) := n - \log\left(\#{\cal C}\right)
\end{equation}
where logarithms without any subscript are to the base two throughout.






We have the following result.
\begin{Theorem}\label{thm1} For all~\(n \geq 3,\) there is a~\(n-\)length code~\({\cal C}_{ord}\) with redundancy
\begin{equation}\label{rc0}
R({\cal C}_{ord}) \leq \log(n+1) +\log{3},
\end{equation}
capable of correcting all ordered deletion-erasure patterns. Conversely, if~\({\cal D}\) is any~\(n-\)length code
capable of correcting all ordered deletion-erasure patterns, then the redundancy
\begin{equation}\label{del_er_red}
R({\cal D}) \geq \log\left(n-1 - 2\sqrt{(n-1)\log{n}}\right)
\end{equation}
for all~\(n\) large. 
\end{Theorem}
We use the single deletion correcting VT codes with minor modifications to construct the code in Theorem~\ref{thm1}. The proof of Theorem~\ref{thm1} in Section~\ref{pf} also contains the decoding algorithm for correcting the ordered deletion-erasure. The lower bound~(\ref{del_er_red}) is true since any code~\({\cal D}\) capable of correcting all ordered deletion-erasure patterns must also be capable of correcting a single deletion. Therefore~(\ref{del_er_red}) follows from the redundancy lower bound for single deletion correcting code in Levenshtein (1965), Sloane (2000). For completeness we provide a small proof using deviation estimates in Section~\ref{pf_red_min}.

The paper is organized as follows: In Section~\ref{pf}, we provide the encoding and decoding in Theorem~\ref{thm1} and obtain the redundancy upper bound~(\ref{rc0}). In Section~\ref{pf_red_min}, we prove the redundancy lower bound~(\ref{del_er_red}) and in Section~\ref{sim_res},
we plot and describe the bounds obtained in Theorem~\ref{thm1} and state our conclusion.

\renewcommand{\theequation}{\arabic{section}.\arabic{equation}}
\renewcommand{\theLemma}{\arabic{section}.\arabic{Lemma}}
\setcounter{equation}{0}
\section{Encoding and Decoding in Theorem~\ref{thm1}} \label{pf}
For integers~\(n \geq 3\) and~\(0 \leq a_1\leq 2\) and~\(0 \leq a_2 \leq n,\) define the two parameter Varshamov-Tenengolts (VT) code~\(VT_{a_1,a_2}(n)\) to be the set of all vectors~\(\mathbf{x}  = (x_1,\ldots,x_n) \in \{0,1\}^n\) such that
\begin{equation}\label{chck_sum}
\sum_{i=1}^{n} x_i \equiv a_1 \mod{3} \text{ and } \sum_{i=1}^{n} i \cdot x_i \equiv a_2 \mod(n+1).
\end{equation}
We show below that~\(VT_{a_1,a_2}(n)\) is capable of correcting all ordered deletion-erasure patterns.

We first derive the decoding algorithm for correcting words corrupted by an ordered deletion-erasure pattern consisting of a single deletion and a single erasure. Later, we show how the algorithm can also be used for words corrupted by a single deletion alone. Suppose~\(\mathbf{x} \in VT_{a_1,a_2}(n)\) is the word transmitted and~both deletion and erasure occur. Given~\(\mathbf{y} = F_{d,e}(\mathbf{x})\) with~\(1 \leq d \leq e \leq n-1\) the location of the erasure~\(e\) is known and we are interested in determining where deletion has occurred. We use an auxiliary parameter~\(k\) that is a candidate for the deletion index~\(d\) and compute checksums analogous to~(\ref{chck_sum}) by varying~\(k\) until we obtain a synchronization. 


For~\(1 \leq k \leq e\) and~\(\hat{x}_d ,\hat{x}_{e+1} \in \{0,1\},\) define the modified checksum
\begin{equation}\label{ewq1}
f_k(\mathbf{y}) := \sum_{i=1}^{k-1}i \cdot y_i + k \cdot \hat{x}_d + \sum_{i=k, i \neq e}^{n-1}(i+1) \cdot y_i + (e+1)\cdot \hat{x}_{e+1}.
\end{equation}
The proof of Theorem~\ref{thm1} follows if we show that the difference~\(f_k(\mathbf{y}) -a_2 \equiv 0 \mod{n+1} \) if and only if the following two conditions are satisfied:\\
\((q1)\) The bits~\(\hat{x}_d = x_d\) and~\(\hat{x}_{e+1} = x_{e+1}\) and\\
\((q2)\) The index~\(k\) belongs to the run containing the deleted bit; i.e., \(d_1 \leq k \leq d_2\) where~\(d_1\) and~\(d_2\) are such that~\(x_{k} =x_j\) for all~\(d_1 \leq k, j \leq d_2\) and~\(x_{d_1-1} = x_{d_2+1} = 1-x_{d_1}.\)

We do so by evaluating~\(f_k(.)\) below case by case. Throughout, if~\(k_1 < k_2\) then~\(\sum_{i=k_2}^{k_1}\) is interpreted as zero.\\\\
\emph{Case I}:~\(1 \leq k \leq d.\)\\
Since~\(y_i = x_i\) for~\(1 \leq i \leq d-1\) and~\(y_i = x_{i+1}\) for~\(d \leq i \leq n-1, i \neq e,\) we get
that
\begin{equation}\label{ewq33}
\sum_{i=1}^{k-1} i \cdot y_i = \sum_{i=1}^{k-1} i \cdot x_i
\end{equation}
and that~\(\sum_{i=k, i \neq e}^{n-1}(i+1) \cdot y_i\) equals
\begin{eqnarray}
&&\sum_{i=k}^{d-1}(i+1)\cdot x_i + \sum_{i=d, i \neq e}^{n-1} (i+1) \cdot x_{i+1} \nonumber\\
&&=\sum_{i=k}^{d-1}i\cdot x_i + \sum_{i=d, i \neq e}^{n-1} (i+1) \cdot x_{i+1}  + \sum_{i=k}^{d-1} x_i\nonumber\\
&&=\sum_{i=k}^{d-1}i\cdot x_i + \sum_{i=d}^{n-1} (i+1) \cdot x_{i+1}  \nonumber\\
&&\;\;\;\;\;\;\;\;\;\;\;\;+ \sum_{i=k}^{d-1} x_i - (e+1) \cdot x_{e+1}\nonumber\\
&&=\sum_{i=k}^{n}i\cdot x_i + \sum_{i=k}^{d-1} x_i - (e+1) \cdot x_{e+1} - d \cdot x_d. \label{ewq3}
\end{eqnarray}
Substituting~(\ref{ewq3}) and~(\ref{ewq33}) into~(\ref{ewq1}) we then get
\begin{equation}\label{fk_eval11}
f_k(\mathbf{y}) =  \Delta + r(k,d-1) + \sum_{i=1}^{n}i \cdot x_i,
\end{equation}
where
\begin{equation}\label{del_def}
\Delta := k \cdot \hat{x}_d -d \cdot x_d + (e+1)\cdot \hat{x}_{e+1} - (e+1) \cdot x_{e+1}
\end{equation}
and
\begin{equation}\label{r_def}
r(k,d-1) := \sum_{i=k}^{d-1} x_i.
\end{equation}
Using the check sum condition~(\ref{chck_sum}) we then get
\begin{equation}\label{fk_eval1}
f_k(\mathbf{y}) - a_2 \equiv \Delta + r(k,d-1) \mod (n+1).
\end{equation}

\emph{Case II}:~\(d+1 \leq k \leq e.\) 
We split the first sum in~(\ref{ewq1}) as
\begin{eqnarray}
\sum_{i=1}^{k-1}i \cdot y_i &=& \sum_{i=1}^{d-1}i \cdot y_i + \sum_{i=d}^{k-1} i \cdot y_i \nonumber\\
&=& \sum_{i=1}^{d-1} i \cdot x_i + \sum_{i=d}^{k-1} i \cdot x_{i+1} \nonumber\\
&=& \sum_{i=1}^{k} i \cdot x_i - d \cdot x_d - \sum_{i=d}^{k-1}  x_{i+1} \nonumber\\
&=& \sum_{i=1}^{k} i \cdot x_i - d \cdot x_d - r(d+1,k),\label{case2_ew1}
\end{eqnarray}
by~(\ref{r_def}).

The second sum in~(\ref{ewq1}) is
\begin{eqnarray}
\sum_{i=k, i \neq e}^{n-1}(i+1) \cdot y_i &=& \sum_{i=k, i \neq e}^{n-1}(i+1)\cdot x_{i+1} \nonumber\\
&=&\sum_{i=k}^{n-1}(i+1)\cdot x_{i+1} - (e+1) \cdot x_{e+1}. \nonumber\\
\label{case2_ew2}
\end{eqnarray}
Substituting~(\ref{case2_ew1}) and~(\ref{case2_ew2}) into~(\ref{ewq1}) we get
\begin{equation}\label{fk_eval22}
f_k(\mathbf{y}) =  \Delta - r(d+1,k) + \sum_{i=1}^{n}i \cdot x_i
\end{equation}
and again using~(\ref{chck_sum}) we get
\begin{equation}\label{fk_eval2}
f_k(\mathbf{y}) - a_2 \equiv \Delta - r(d+1,k) \mod(n+1).
\end{equation}

To proceed with the evaluation, we use the first relation in~(\ref{chck_sum}) to get that
\begin{equation}\label{chk1}
D(\mathbf{y}) := \left(a_1 - \sum_{i=1}^{n-1} y_i\right) \mod{3} \in \{0,1,2\}.
\end{equation}
We consider three cases separately.\\
\emph{Case~\((a)\)}:~\(D(\mathbf{y}) = 0.\)\\
In this case, the deleted bit~\(x_d\) and the erased bit~\(x_{e+1}\) of~\(\mathbf{x}\) are both zero and we set~\(\hat{x}_{e+1} = \hat{x}_d = 0.\) Thus~\(\Delta\) defined in~(\ref{del_def}) is zero. Suppose that~\(d_1 \leq d \leq d_2\) are such that~\(x_{i} = 0\) for~\(d_1 \leq d \leq d_2\) and~\(x_{d_1-1} = x_{d_2+1} = 1.\) In other words,~\((x_{d_1},\ldots,x_{d_2})\) is the run of zeros containing the deleted bit~\(x_d=0.\) If~\(1 \leq k \leq d_1-1,\) then there is at least one index between~\(k\) and~\(d_1-1\) whose bit value is one and so~\(1 \leq r(k,d-1) \leq n.\) Therefore we get from~(\ref{fk_eval1}) that~\(f_k(\mathbf{y}) - a_2 \neq 0  \mod(n+1).\) If~\(d_1 \leq k \leq d\) then~\(r(k,d-1) = 0\) and if~\(d+1 \leq k \leq d_2\) then~\(r(d+1,k) = 0.\) Therefore from~(\ref{fk_eval1}) and~(\ref{fk_eval2}) we get that~\(f_k(\mathbf{y})-a_2 \equiv 0 \mod(n+1)\) for~\(d_1 \leq k \leq d_2.\) Finally if~\(d_2+1 \leq k \leq e,\) then again \(-n \leq -r(d+1,k) \leq -1\) and so from~(\ref{fk_eval2}) we again have that~\(f_k(\mathbf{y}) - a_2 \neq 0 \mod(n+1).\)

\emph{Case~\((b)\)}:~\(D(\mathbf{y}) = 2.\)\\
In this case, the deleted bit~\(x_d\) and the erased bit~\(x_{e+1}\) of~\(\mathbf{x}\) are both one and we set~\(\hat{x}_{e+1} = \hat{x}_d = 1.\) Thus the term~\(\Delta\) defined in~(\ref{del_def}) then equals~\(k-d.\) In this case, we let~\(d_1 \leq d \leq d_2\) be such that~\(x_{i} = 1\) for~\(d_1 \leq d \leq d_2\) and~\(x_{d_1-1} = x_{d_2+1} = 0.\) If~\(k \leq d_1-1,\) then there is at least one zero between the indices~\(k\) and~\(d-1\) and so
we get from~(\ref{fk_eval1}) that
\begin{equation}\label{gr1}
-n \leq  k-d+r(k,d-1)  = -\sum_{i=k}^{d-1}(1-x_i) \leq -1.
\end{equation}
Therefore from~(\ref{fk_eval1})~\(f_k(\mathbf{y}) - a_2 \neq 0 \mod(n+1).\) If~\(d_1 \leq k \leq d\) then using~(\ref{gr1}), we have that~\(k-d+r(k,d-1)=0\)
and if~\(d+1 \leq k \leq d_2,\) then~\[k-d - r(d+1,k)  = \sum_{i=d+1}^{k}(1-x_i) = 0.\] From~(\ref{fk_eval1}) and~(\ref{fk_eval2}) we therefore have that~\(f_k(\mathbf{y})-a_2 \equiv 0 \mod(n+1)\) for~\(d_1 \leq k \leq d_2.\) Finally if~\(d_2+1 \leq k \leq e,\) then there is at least one zero between the indices~\(d+1\) and~\(k\) and so from~(\ref{fk_eval2}),
\begin{equation}\label{gr2}
n \geq k-d - r(d+1,k)  = \sum_{i=d+1}^{k}(1-x_i)  \geq 1.
\end{equation}
From~(\ref{fk_eval2}), we then get that~\(f_k(\mathbf{y}) - a_2 \neq 0 \mod(n+1).\)

\emph{Case~\((c)\)}:~\(D(\mathbf{y}) = 1.\)\\
We first suppose that the deleted bit~\(x_d = 1\) and the erased bit~\(x_{e+1} = 0.\) We consider two subcases by setting~\(\hat{x}_{e+1} = 1-\hat{x}_d = 0\) and~\(\hat{x}_{e+1} = 1-\hat{x}_d=1.\)

\emph{Subcase~\((c1)\)}:~\(\hat{x}_{e+1} = 1-\hat{x}_d = 0.\) In this case, the term~\(\Delta\) defined in~(\ref{del_def}) equals~\(k-d.\)
Arguing as in case~\((b)\) we get that~\(f_k(\mathbf{y}) -a_2 \equiv 0 \mod(n+1)\) if and only if~\(d_1 \leq k \leq d_2\) i.e.,~\(k\)
belongs to the run containing the deleted bit~\(x_d.\)

\emph{Subcase~\((c2)\)}:~\(\hat{x}_{e+1} = 1-\hat{x}_d = 1.\) Here~\(\Delta=-d+e+1\)
and if~\(1 \leq k \leq d,\) then we use the estimate
\begin{equation}\label{rkd_est}
r(a,b) =\sum_{i=a}^{b} x_i \leq b-a+1
\end{equation}
with~\(a = k\) and~\(b = d-1\) to get that
\begin{equation}\label{eq_c22}
-k+e+1 \geq -d+e+1 + r(k,d-1) \geq -d+e+1 \geq 1
\end{equation}
since~\(d \leq e.\) Using~\(k \leq e \leq n\) we therefore get from~(\ref{fk_eval1}) and~(\ref{eq_c22}) that the difference~\(f_k(\mathbf{y}) - a_2 \neq 0 \mod(n+1).\)

If~\(d+1 \leq k \leq e,\) then we use~(\ref{rkd_est}) with~\(a = d+1\) and~\(b = k\) to get that
\begin{eqnarray}
e&\geq&-d+e+1 - r(d+1,k) \nonumber\\
&\geq& -d+e+1 -(k-d) \nonumber\\
&=& e+1-k \nonumber\\
&\geq& 1, \nonumber
\end{eqnarray}
since~\(k \leq e.\)  Using~(\ref{fk_eval2}) and~\(e \leq n,\) we again have that~\(f_k(\mathbf{y}) - a_2 \neq 0 \mod(n+1).\)

Suppose now that the deleted bit~\(x_d = 0\) and the erased bit~\(x_{e+1} = 1.\) We again consider two subcases by setting~\(\hat{x}_{e+1} = 1-\hat{x}_d = 0\) and~\(\hat{x}_{e+1} = 1-\hat{x}_d=1.\)

\emph{Subcase~\((c1)\)}:~\(\hat{x}_{e+1} = 1-\hat{x}_d = 0.\)
Here~\(\Delta=k-e-1\) and if~\(1 \leq k \leq d,\) then using~(\ref{rkd_est}) with~\(a = k\) and~\(b  =d-1,\) we get that \[-e \leq  k-e-1 + r(k,d-1) \leq k-e-1+(d-k) = d-e-1 \leq -1 \] since~\(d \leq e.\)
Again using~\(e \leq n\) we get from~(\ref{fk_eval1}) that the difference~\(f_k(\mathbf{y}) - a_2 \neq 0 \mod(n+1).\)

If~\(d+1 \leq k \leq e,\) then using~(\ref{rkd_est}) with~\(a = d+1\) and~\(b =k\) we get that
\begin{eqnarray}
-1&\geq&k-e-1 - r(d+1,k) \nonumber\\
&\geq& k-e-1 -(k-d) \nonumber\\
&=& d-e-1 \nonumber\\
&\geq& -e, \nonumber
\end{eqnarray}
since~\(d \geq 1.\)  Using~(\ref{fk_eval2}) and the fact that~\(e \leq n,\) we again have that~\(f_k(\mathbf{y}) - a_2 \neq 0 \mod(n+1).\)

\emph{Subcase~\((c2)\)}:~\(\hat{x}_{e+1} = 1-\hat{x}_d = 1.\) In this case, the term~\(\Delta\) defined in~(\ref{del_def}) is zero.
Arguing as in case~\((a)\) we get that~\(f_k(\mathbf{y}) -a_2 \equiv 0 \mod(n+1)\) if and only if~\(d_1 \leq k \leq d_2\) i.e.,~\(k\)
belongs to the run containing the deleted bit~\(x_d.\)


Finally, we now suppose that~\(e = n\) and that the word~\(\mathbf{y} = F_{d,e}(\mathbf{x})\) is obtained after corruption by a single deletion alone. To
obtain~\(\mathbf{x}\) from~\(\mathbf{y},\) we simply use the above algorithm with~\(e = n\) and~\(\hat{x}_{e+1} = 0.\)
The term~\(\Delta\) defined in~(\ref{del_def}) is then~\(k \cdot \hat{x}_d - d \cdot x_d.\) The auxiliary quantity~\(f_k(\mathbf{y})\)
defined in~(\ref{ewq1}) evaluates to the expression in~(\ref{fk_eval1}) for~\(1 \leq k \leq d\) and
the expression~(\ref{fk_eval2}) for~\(d+1 \leq k \leq e = n.\) The discrepancy~\(D(\mathbf{y})\) defined in~(\ref{chk1})
gives us the value of the deleted bit~\(x_d.\) If~\(x_d =0,\) then~\(\Delta = 0\) and arguing as in  case~\((a)\)
above, we get that~\(f_k(\mathbf{y}) - a_2 \equiv 0 \mod(n+1)\)
if and only if~\(d_1 \leq k \leq d_2,\) the run containing the deleted bit~\(x_d.\)
If~\(x_d = 1,\) then~\(\Delta = k-d\) and we argue as in case~\((b)\) above to determine
the run containing the deleted bit~\(x_d.\)

The above procedure is summarized in Algorithm~\ref{euclid4} below. To compute the redundancy of~\(VT_{a_1,a_2}(n),\) we choose~\(a_1\) and~\(a_2\) such that~\(\#VT_{a_1,a_2}(n) \geq \frac{2^{n}}{3(n+1)}\) so that~\(R(VT_{a_1,a_2}(n)) \leq \log(3(n+1)).\) This proves~(\ref{rc0}). We prove the lower bound on the redundancy~(\ref{del_er_red}), in the next section.~\(\qed\)

\begin{figure}[!t]
 \removelatexerror

\begin{algorithm}[H]
\caption{Correcting an ordered deletion-erasure}\label{euclid4}
\SetNoFillComment

\Input{Received word $\mathbf{y},$ erasure position~$e$ (If no erasure, $e \gets n$)}
\Output{Estimated word $\mathbf{\hat{x}}$}%
\OrdDecode{}
{
\emph{Initialization}: $k \gets 0, CS(0) \gets a_2+1 \mod(n+1)$\;
\emph{Preprocessing}: $D \gets \left(a_1 - \sum_{i=1, i\neq e}^{n-1} y_i \right) \mod {3}$\;

\tcc{Compute estimates of deleted and erased bits}
\eIf{$no\;erasure$}
{$\hat{x}_d \gets D, \hat{x}_{e+1} \gets 0$\; }
{
\eIf{$D \in \{0,2\}$}
{
$\hat{x}_d \gets \frac{D}{2} , \hat{x}_{e+1} \gets \frac{D}{2}$\;
}
{
$\hat{x}_d \gets 1, \hat{x}_{e+1} \gets 0, Flag \gets 0$\;
}
}
\tcc{Pass all positions with non zero discrepancy until erasure}
\emph{Main Loop}:

\While {$CS(k)-a_2 \neq 0 \mod(n+1) $ and $k \leq e$}{
    $k \gets k+1$\;
    $CS(k) \gets \sum_{i=1}^{k-1}i \cdot y_i + k \cdot \hat{x}_d  + \;\;\;\;\;\; \sum_{i=k, i \neq e}^{n-1}(i+1) \cdot y_i + (e+1)\cdot \hat{x}_{e+1}$\;
}

\tcc{check discrepancy}
\If{$CS(k)-a_2 \equiv 0 \mod(n+1)$}
{
\textbf{output}
$\mathbf{\hat{x}} = (y_1,\ldots,y_{k-1},\hat{x}_d,y_{k},\ldots,y_{e-1},\hat{x}_{e+1},y_{e+1},\ldots,y_{n-2})$\;
}
\If{$Flag=1$}
{\textbf{output} ``Error"}

\tcc{Perform loop with revised values}
{

$\hat{x}_d \gets 0, \hat{x}_{e+1} \gets 1, k \gets 0, Flag \gets 1, CS(0) \gets a_2+1 \mod(n+1)$\;
\textbf{goto} \emph{Main Loop}\;

}


}
\end{algorithm}

\end{figure}

\setcounter{equation}{0}
\renewcommand\theequation{\arabic{section}.\arabic{equation}}
\section{Proof of the redundancy lower bound} \label{pf_red_min}
We use the following standard deviation result. Let~\(\{X_j\}_{1 \leq j \leq m}\) be independent
Bernoulli random variables with~\[\mathbb{P}(X_j = 1) = \frac{1}{2} = 1-\mathbb{P}(X_j = 0)\] and
let~\(0 < \epsilon   = \epsilon(m) \leq \frac{1}{2}.\) If~\(T_m = \sum_{j=1}^{m} X_j,\) then
\begin{equation}\label{conc_est_f}
\mathbb{P}\left(\left|T_m - \frac{m}{2}\right| \geq \frac{m \epsilon}{2} \right) \leq 2\exp\left(-\frac{m\epsilon^2}{2}\right)
\end{equation}
for all \(m \geq 1.\)\\
\emph{Proof of~(\ref{conc_est_f})}: Using Corollary A.1.2, pp. 308 of Alon and Spencer (2008) with~\(S_m := \sum_{i=1}^{m} 2X_i-1 = 2T_m-m,\) we get
\begin{equation}\nonumber
\mathbb{P}\left(\left|2T_m - m\right| \geq a \right) \leq 2\exp\left(-\frac{a^2}{2m}\right).
\end{equation}
Setting~\(a = m\epsilon,\) we get~(\ref{conc_est_f}).~\(\qed\)

The following result determines the fraction of words in~\(\{0,1\}^{n}\) containing
a large number of runs. Recall that if~\(\mathbf{x} = (x_1,\ldots,x_n),\)
then~\((x_{k_1},x_{k_1+1},\ldots,x_{k_2})\) is said to be a run if~\(x_{k} = x_j\)
for~\(k_1 \leq k,j \leq k_2\) and~\(x_{k_1-1} \neq x_{k_1}\) and~\(x_{k_2+1} \neq x_{k_2}.\)
For~\(\mathbf{x} \in \{0,1\}^{n}\) let~\(Q(\mathbf{x})\)
denote the number of runs in~\(\mathbf{x}\) and let
\begin{equation}\label{v_def}
{\cal U}_n := \left\{\mathbf{x} \in \{0,1\}^{n} : Q(\mathbf{x}) \geq \frac{n-1}{2} - \sqrt{2(n-1)\log{n}}\right\}.
\end{equation}
We have the following Lemma.
\begin{Lemma}\label{run_lem2} For all~\(n\) large,
\begin{equation}\label{v_est}
\#{\cal U}_n \geq 2^{n}\left(1-\frac{4}{n^2}\right).
\end{equation}
\end{Lemma}
\emph{Proof of Lemma~\ref{run_lem2}}: Let~\(\mathbf{X} = (X_1,\ldots,X_n)\) be a uniformly randomly chosen word in~\(\{0,1\}^{n}\)
so that~\(X_i, 1 \leq i \leq n\) are independent and identically distributed (i.i.d.)
with \[\mathbb{P}(X_1 = 0) = \frac{1}{2} = \mathbb{P}(X_1 = 1).\]

The number of runs
\begin{equation}\label{q_split}
Q(\mathbf{X}) = 1 + \sum_{i=1}^{n-1} \ind(X_i \neq X_{i+1}) = 1+Q_1+Q_2,
\end{equation}
where~\(Q_1 = \sum_{i=1, i\;\text{even}}^{n-1}  \ind(X_i \neq X_{i+1})\) and\\\(Q_2 = \sum_{i=1, i\;\text{odd}}^{n-1}  \ind(X_i \neq X_{i+1}).\)
The term~\(Q_1\) is a sum consisting of~\(m \geq \frac{n-1}{2}\) i.i.d. random Bernoulli variables \[Z_i = \ind(X_i \neq X_{i+1})\]
with~\(\mathbf{P}(Z_i = 1) = \frac{1}{2} = \mathbf{P}(Z_i = 0)\) and so using~(\ref{conc_est_f})
we get
\begin{equation}\nonumber
\mathbb{P}\left(Q_1 \geq \frac{n-1}{4}(1-\epsilon) \right) \geq 1-2\exp\left(-\frac{(n-1)\epsilon^2}{4}\right).
\end{equation}
An analogous estimate holds for~\(Q_2\)
and so from~(\ref{q_split}) we get
\begin{equation}\label{t1_est_2}
\mathbb{P}\left(Q(\mathbf{X}) \geq \frac{n-1}{2}(1-\epsilon) \right) \geq 1-4\exp\left(-\frac{(n-1)\epsilon^2}{4}\right).
\end{equation}
Letting~\(\log_e{n}\) be the natural logarithm of~\(n\) and setting~\(\epsilon = \sqrt{\frac{8\log_e{n}}{n-1}} \leq \sqrt{\frac{8\log{n}}{n-1}}\) in~(\ref{t1_est_2}) we then get \[\mathbb{P}\left(Q(\mathbf{X}) \geq \frac{n-1}{2}-\sqrt{2(n-1)\log{n}}\right) \geq 1-\frac{4}{n^2},\] proving~(\ref{v_est}).~\(\qed\)

We use Lemma~\ref{run_lem2} to prove the redundancy lower bound in Theorem~\ref{thm1}.\\
\emph{Proof of~(\ref{del_er_red}) in Theorem~\ref{thm1}}:
Let~\({\cal D}\) be any code capable of correcting all ordered deletion-erasure patterns.
The code~\({\cal D}\) is capable of correcting a single deletion and
we therefore henceforth consider only deletions.
For~\(\mathbf{x} \in {\cal D},\)
let~\({\cal N}(\mathbf{x}) = \cup_{1 \leq d \leq n}\{F_{d,n}(\mathbf{x})\} \subseteq \{0,1\}^{n-1}\) be the set of all possible corrupted words obtained
from~\(\mathbf{x}\) after a single deletion alone. 

By definition, if~\(\mathbf{x}_1 \neq \mathbf{x}_2 \in {\cal D},\) then necessarily
\[{\cal N}(\mathbf{x}_1) \bigcap {\cal N}(\mathbf{x}_2) = \emptyset,\] because
otherwise~\({\cal D}\) would not be capable of correcting a single deletion
and therefore
\begin{equation}\label{sum_less11}
\sum_{\mathbf{x} \in {\cal D}} \#{\cal N}(\mathbf{x}) \leq 2^{n-1}.
\end{equation}
Letting~\({\cal U}_n \) be the set as defined in~(\ref{v_def}) we obtain a lower bound
on~\(\#{\cal N}(\mathbf{x})\) for each word~\(\mathbf{x} \in {\cal D} \bigcap {\cal U}_n.\)

If~\(\mathbf{x} \in {\cal U}_n,\) then there are at least \[\omega_n := \frac{n-1}{2}-\sqrt{2(n-1)\log{n}}\] runs in~\(\mathbf{x}.\)
Deleting one bit in each such run, we get a set of distinct corrupted
words and so~\(\#{\cal N}(\mathbf{x}) \geq \omega_n.\)
From~(\ref{sum_less11}) we therefore get
\begin{equation}\nonumber
\omega_n \cdot \#\left({\cal D} \bigcap {\cal U}_n\right) \leq \sum_{\mathbf{x} \in {\cal D} \bigcap {\cal U}_n} {\cal N}(\mathbf{x}) \leq 2^{n-1}
\end{equation}
and so using~(\ref{v_est}) we get
\begin{eqnarray}
\#{\cal D} &\leq& \frac{2^{n-1}}{\omega_n} + \#{\cal U}^c_n \nonumber\\
&\leq& \frac{2^{n-1}}{\omega_n} + \frac{4\cdot 2^{n}}{n^2} \nonumber\\
&=& \frac{2^{n}}{n-1-\sqrt{8(n-1)\log{n}}} + \frac{4\cdot 2^{n}}{n^2} \nonumber\\
&\leq& \frac{2^{n}}{n-1-2\sqrt{(n-1)\log{n}}} \nonumber
\end{eqnarray}
for all~\(n\) large. This proves~(\ref{del_er_red}).~\(\qed\)


\section{Simulation and Conclusion}\label{sim_res}
\begin{figure}[tbp]
\centering
\includegraphics[width=3.5in]{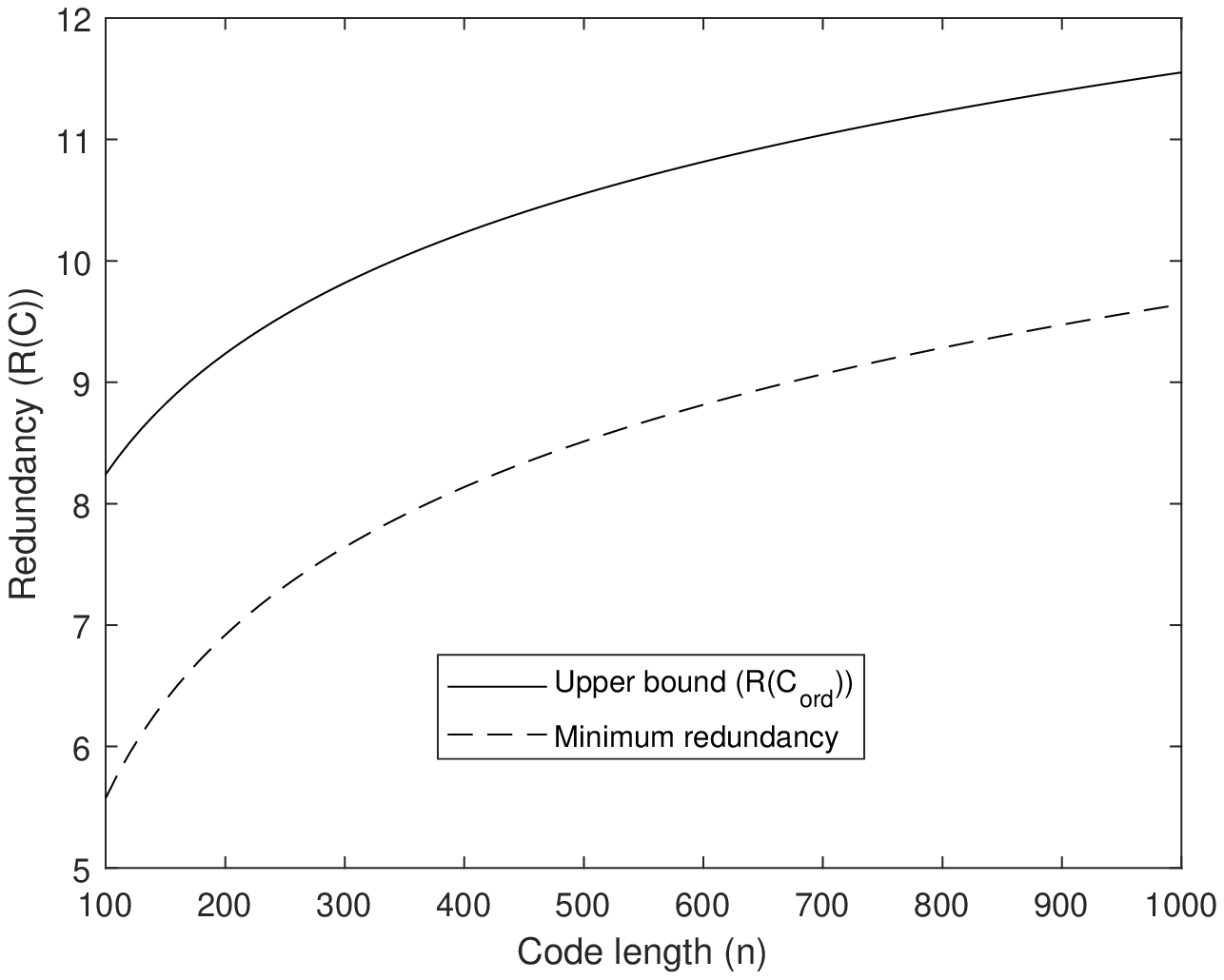}
\caption{Redundancy bounds from Theorem~\ref{thm1}.}
\label{rect_comp}
\end{figure}

In Figure~\ref{rect_comp}, we have plotted the redundancy bounds obtained in Theorem~\ref{thm1}.
The solid line corresponds to the upper bound~(\ref{rc0}) on the redundancy of the code~\({\cal C}_{ord}.\)
The dotted line is lower bound~(\ref{del_er_red}) on the minimum redundancy of any code
capable of correcting all ordered deletion-erasure patterns. Asymptotically,
the difference converges to~\(\log{3},\) the maximum extra information needed to
correct ordered deletion-erasures.

In this paper, we have shown that the single deletion
correcting VT code with minor modifications can also
be used to correct ordered deletion-erasure patterns
where the deletion always occurs before the erasure.
Moreover, the modified code has the same
logarithmic redundancy as optimal codes.


\section*{Acknowledgements}
I thank Professors Alberto Gandolfi and Federico Camia for crucial comments and for my fellowships.



\setcounter{equation}{0} \setcounter{Lemma}{0} \renewcommand{\theLemma}{II.%
\arabic{Lemma}} \renewcommand{\theequation}{II.\arabic{equation}} %
\setlength{\parindent}{0pt}




%





\bibliographystyle{plain}

\end{document}